\documentclass[5p,twocolumn]{elsarticle}

\usepackage{graphicx}
\usepackage{pslatex}
\usepackage{xspace}
\usepackage{subfigure}
\usepackage{ctable}

\newcommand{\pht}{poly(3-hexyl thiophene)\xspace}
\newcommand{\pcbm}{[6,6]-phenyl-C$_{61}$ butyric acid methyl ester\xspace} 
\newcommand{\ppv}{poly(phenylene vinylene)\xspace}
\newcommand{\degree}{$^{\circ}$}

\newcommand{\pdi}{perylene-3,4:9,10-bis(dicarboximide)\xspace}
\newcommand{\tdi}{terrylene-3,4:11,12-bis(dicarboximide)\xspace} 

\begin{document}

\title{Detailed study of N,N'--(diisopropylphenyl)-- \tdi as electron acceptor for solar cells application.}

\author[ep6]{J.~Gorenflot \corref{cor1}}\ead{gorenflot@physik.uni-wuerzburg.de}

\author[ep6]{A.~Sperlich}

\author[ep6]{A.~Baumann}

\author[zae]{D.~Rauh}

\author[mpip]{A.~Vasilev}

\author[mpip]{C.~Li}

\author[mpip]{M.~Baumgarten}

\author[ep6]{C.~Deibel}

\author[ep6,zae]{V.~Dyakonov \corref{cor1}}\ead{dyakonov@physik.uni-wuerzburg.de}

\cortext[cor1]{Corresponding authors}

\address[ep6]{Experimental Physics VI, Julius-Maximilians-University of W{\"u}rzburg, 97074 W{\"u}rzburg, Germany}
\address[mpip]{Max-Planck Institute for Polymer Research, Ackermannweg 10, 55128 Mainz, Germany}
\address[zae]{Bavarian Centre for Applied Energy Research (ZAE Bayern), 97074 W{\"u}rzburg, Germany}

\date{\today}

\begin{abstract}
We report on \tdi  (TDI) as electron acceptor for bulk-heterojunction solar cells  using \pht (P3HT) as complementary donor component. Enhanced absorption was observed in the blend compared to pure P3HT. As shown by the very efficient photoluminescence (PL) quenching, the generated excitons are collected at the interface  between the donor and acceptor, where they separate into charges which we detect by photoinduced absorption and electron-spin resonance (ESR). Time-of-flight (TOF) photoconductivity measurements reveal a good electron mobility of $10^{-3}$ cm$^2$V$^{-1}$s$^{-1}$ in the blend. Nevertheless, the photocurrent in solar cells was found to be surprisingly low. Supported by the external quantum efficiency (EQE) spectrum as well as morphological studies by way of X-ray diffraction and atomic force microscopy, we explain our observation by the formation of a TDI hole blocking layer at the anode interface which prevents the efficiently generated charges to be extracted.
\end{abstract}


\maketitle

Bulk heterojunction solar cells belong to the most promising solutions to produce cheaper solar energy due to their low-cost solution processability.\cite{chopra2004review, deibel2010review3} Nevertheless, their efficiency and long-term stability require improvement. Moreover the complex mechanisms underlying the photovoltaic effect in those intrinsically disordered materials are still poorly understood. For those reasons, the quest for new photoactive materials is still a topical issue. Furthermore, the study of systems differing from the reference ones is also a necessity to understand the specific properties that enable those reference systems to reach relatively good efficiencies.

Since the first attempts of using conductive polymers to transform solar energy into electricity, \cite{tani1980} many obstacles have already been overcome. The use of electron donor:acceptor heterojunctions enabled to split the primary photoexcitation, where charges are strongly bound to each other, into separated polarons. The difficulty of solution-processing bilayers was avoided by the use of self-organising blends --- so called bulk heterojunctions, such as \ppv : \pcbm (PPV:PCBM) --- enabling for the first time in the 90's to reach the percent efficiency with solution processed devices.\cite{yu1995} And the tendency went on in the 2000's with the emergence of more stable polymers with an important crystalinity such as \pht, whose blend with \pcbm (PCBM) is still a reference composite.\cite{deibel2010review}

Additionally to the two afore-mentioned materials, a variety of polymer electron donors have been used to produce high performance solar cells.\cite{mcneill2007, shin2007a, park2009} In contrast, only a few efficient alternatives to fullerene derivatives as electron acceptor have been reported. Those alternatives being mainly inorganic ---such as ZnO nanocrystals \cite{oosterhout2009}--- or polymeric \cite{mcneill2008}, although conjugated molecules such as \pdi have been long known for their good electron accepting properties.\cite{horowitz1996,jen1999}

\begin{figure}[t]
	\begin{center}
	\includegraphics[width=8cm]{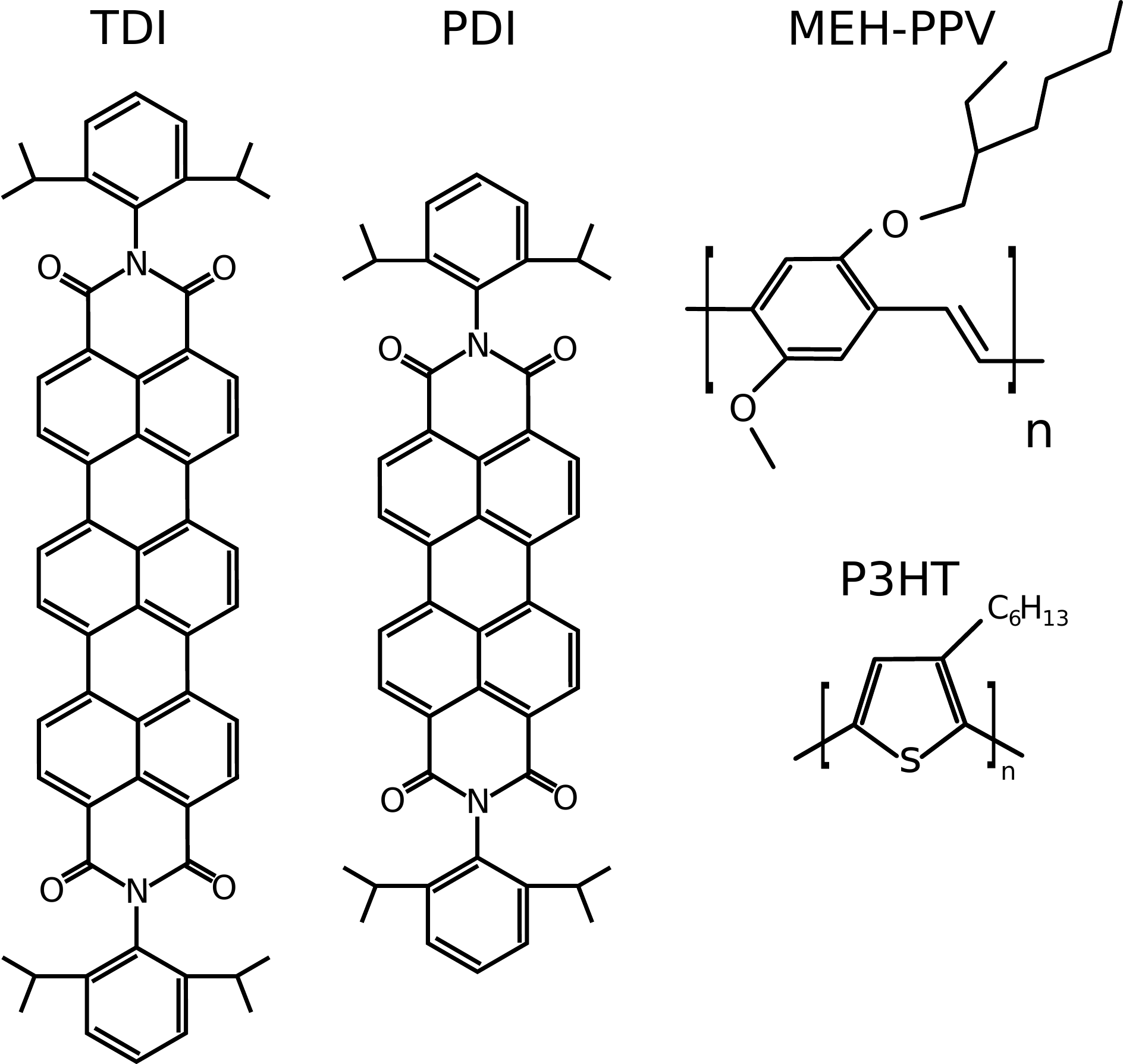}%
	\caption{Chemical structure of the materials. %
	\label{fig:mol}}
	\end{center}
\end{figure}

Why fullerene based cells are performing better than the ones using other acceptors is still an open question. In order to understand better this singularity, a molecule of the rylene-bis(dicarboximide) family: a terrylene diimide (TDI) was studied extensively in blend with P3HT. For comparison purposes we carried out additional studies on blends of perylene diimide (PDI) with P3HT as well as blends of TDI with poly[2-methoxy-5-(2-ethylhexyloxy)-p-phenylene vinylene] (MEH-PPV). The molecules of PDI and TDI used in this work are represented in Fig.~\ref{fig:mol}. Both have electron affinities --- related to their lowest unoccupied molecular orbital (LUMO) --- of 3.7~eV (determined from cyclovoltametry spectra presented by Lee et a.~\cite{lee1999a} using the method presented by Pron et al.~\cite{pron2010review}), making them potential acceptors for higher energy electrons from P3HT (LUMO~=~-~3.3~eV). Their synthesis has already been presented elsewhere.~\cite{nolde2006}

	The photocurrent generation in bulk heterojunctions is a complex mechanism involving several steps. Each of them requires separated study to reach a proper understanding of the material's behavior.\cite{deibel2010review} (a)~Light absorption generates excitons that need to (b)~diffuse to the heterojunction interface where they are (c)~dissociated into charges separated between the two phases. Those charges have to be subsequently (d)~transported through their respective material to finally (e)~reach an electrode and be extracted. Those steps were systematically monitored using the following methods: (a)~thin film optical absorption measurements, (b)~quenching of the excitons photoluminescence, (c)~photo-induced absorption and electron-spin resonance spectroscopies, (d)~time-of-flight transport measurements. Finally, solar cells were built, providing indication about step~(e) by measuring current voltage characteristics and external quantum efficiency spectra. Furthermore the morphology of the blend was investigated by scanning electron microscopy for the surface and X-ray diffraction for the bulk.
		
\section{Experimental methods}

The rylenes used were synthesized by a method presented previously.\cite{nolde2005} We used P3HT 4002E from Rieke Metals which has a regioregularity between 90 and 93 $\%$. MEH-PPV was purchased from Sigma Aldrich. PC$_{60}$BM used for reference P3HT:PCBM blends was purchased from Solenne. Except otherwise stated, samples were processed from 15~mg/ml solutions in chloroform.

Films for optical measurements (absorption, photo-induced absorption and photoluminescence) were spincoated at 1000~rpm for 20~seconds on sapphire substrates. Remaining excess dissolved material and solvent were spinned away by a second 10~s step at 3000~rpm. The samples were then placed on a hot plate at 120~\degree C for 10~minutes annealing. This treatment enables the P3HT self-organisation \cite{cugola2006} and was also applied to samples without P3HT for comparability.

UV-vis absorption spectra were recorded with a Perkin Elmer Lambda 900 spectrometer at room temperature. X-ray diffraction measurements were carried out using a Philips PW 1729 X-ray generator. Scanning electron microscopy (SEM) measurements were carried out using a ZEISS ULTRA plus scanning electron microscope.

The photoinduced absoption spectra were recorded using a photomodulation technique: a transmission spectrum of the sample was realized using a Xenon lamp and a monochromator while the sample is photo-excited by a chopped cw laser. As excitation source, we used the 514~nm emission of a Melles Griot Ar$^+$ cw laser at a power after mechanical chopping of 30~mW. Excitation and probe lights were focussed onto the same point of the sample. The transmitted light was collected by large diameter concave mirrors and focussed into the entrance slit of a cornerstone monochromator. The use of 2 photodiodes (silicon diode from 550 to 1030~nm and liquid nitrogen-cooled InSb diode from 1030~nm to 5500~nm) enabled us to access a wide range of energies from 0.23 to 2.25~eV (550 to 5500~nm). The signals were recorded with a standard phase sensitive technique synchronised with the chopping frequency of the laser by using Signal Recovery 7265 DSP Lock-In amplifier. The chopping frequency used was 330~Hz for spectral studies, low enough to see most of the signal and high enough to gain a good signal to noise ratio. The spectrum is then calculated with equation~\ref{eqn:PIA}.
\begin{equation}
	\frac{-\Delta T}{T}(\lambda) = \frac{PIAPL_x(\lambda) - PL_x(\lambda)}{T(\lambda)}
	\label{eqn:PIA}
\end{equation} 
 \textit{$PIAPL_x$} is the in-phase value measured by the lock-in, \textit{$PL_x$} is the in-phase value under photoluminescence conditions (without probe light) and \textit{T} is the transmitted signal without excitation (by modulating the white light). The phase is set to zero under luminescence conditions because the extremely short lifetime of fluorescing species prevents lifetime related phase shift, thus enabling us to know and compensate the phase shift caused by the measuring system alone. Photoluminescence measurements were carried out using the same setup as for photo-induced absorption. 

ESR (modified Bruker 200D) was applied to verify the presence of spin carrying polarons. The sample was placed in a resonant cavity and cooled with a continuous flow helium cryostat, allowing a temperature range from 10~K to room temperature. The microwave absorption was measured by using lock-in, with modulation of the external magnetic field as reference. For the excitation, a white light halogen lamp guided to the microwave cavity was used. The g-factor of ESR signals was calibrated for every measurement with a Bruker 035 M NMR-Gaussmeter and an EIP 28b frequency counter. Details about the technique can be found elsewhere. \cite{sperlich2011,marumoto2002} The ESR samples were dropcast films from 200~$\mu l$ solution prepared inside a nitrogen glovebox. The resulting films were rolled up and put inside a ESR sample tubes. All samples were annealed afterwards for 10~min at 120~\degree C.

We processed solar cells by depositing the active layer on Poly(3,4-ethylenedioxythiophene) poly(styrenesulfonate) (PEDOT:PSS) covered indium tin oxide (ITO)/glass substrates. The active layer was spin-coated as afore-mentioned. The evaporated cathode was a Ca/Al (3/100~nm) for the solar cells and the active area was 0.5 cm$^2$. For current--voltage characterization we used a 300~W Xenon lamp for illumination. The intensity was adjusted to 100~mWcm$^{-2}$.

In TOF measurements we used a solar-cell like structure with a thicker active layer. The thickness of the films combined with the high absorption coefficient of the material used enables that a pulse of light generates excitons in the neighborhood of one of the electrodes and not in the bulk. The combined action of the field and the electrode separate charges and extract one type immediately after excitation whereas the other type of carrier drifts towards other electrode and reaches it after a transit time which can be related to the mobility by the following equation:
\begin{equation}
	\mu = \frac{d}{Et_{tr}}
	\label{eqn:TOF}
\end{equation}
where \textit{$t_{tr}$} is the transit time, \textit{d} the thickness, \textit{E} the applied field, and \textit{$\mu$} the mobility of the studied charges. We processed films with thickness of around 2~$\mu$m resulting in applied fields in the order of 10$^7$~Vm$^{-1}$. More detail about this experiment can be found elsewhere.\cite{baumann2008}

\section{Experimental results}

\begin{figure}
	\begin{center}
	\includegraphics[width=9cm]{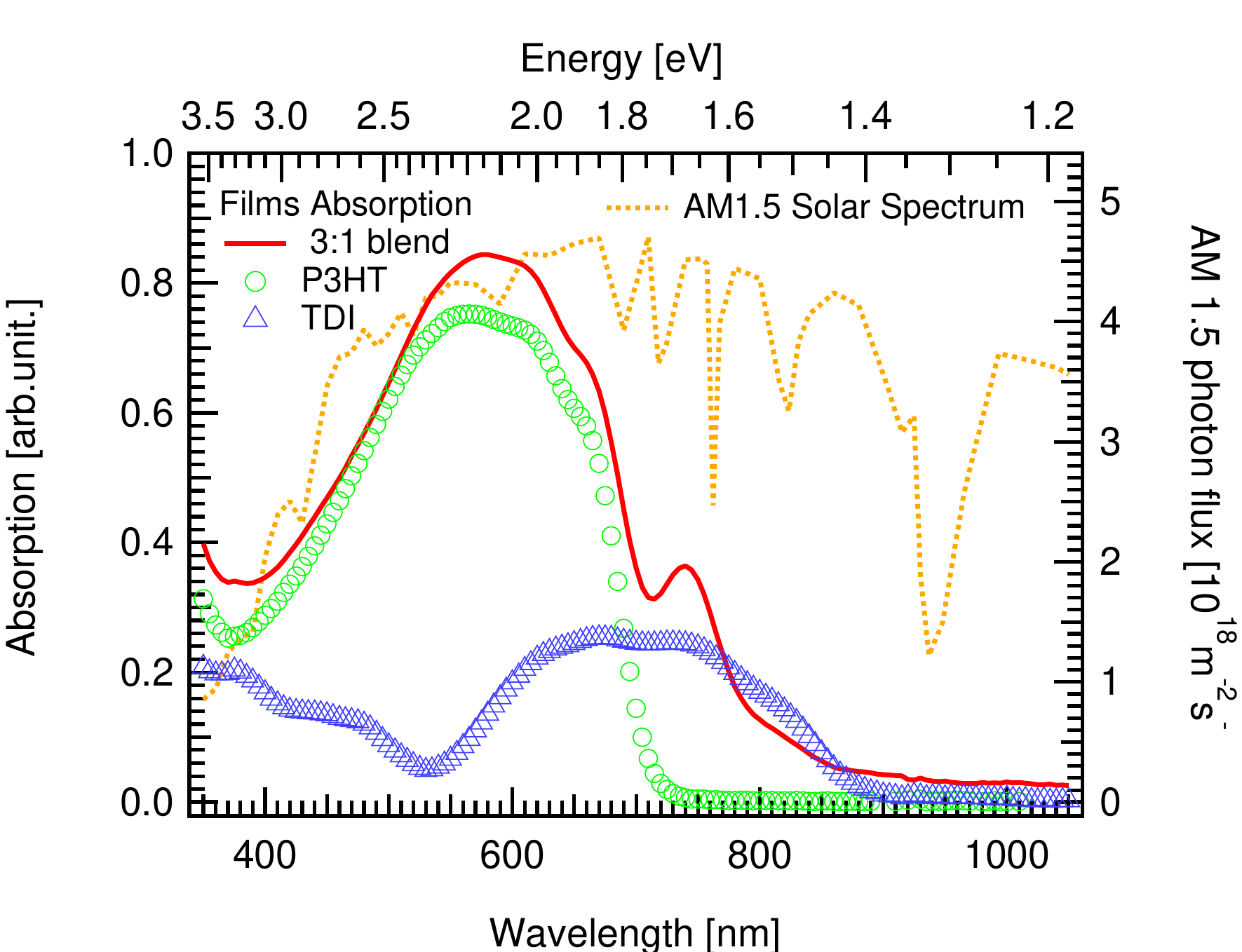}
	\caption{Absorption spectra of pristine P3HT thin film ($\bigcirc$), pristine TDI thin film ($\bigtriangleup$) and 3:1 P3HT:TDI blend (continuous line) together with the AM1.5 solar spectrum (dotted line, right axis).
	\label{fig:absorption}}
	\end{center}
\end{figure}

\begin{figure}[t]
	\begin{center}
	\includegraphics[width=9cm]{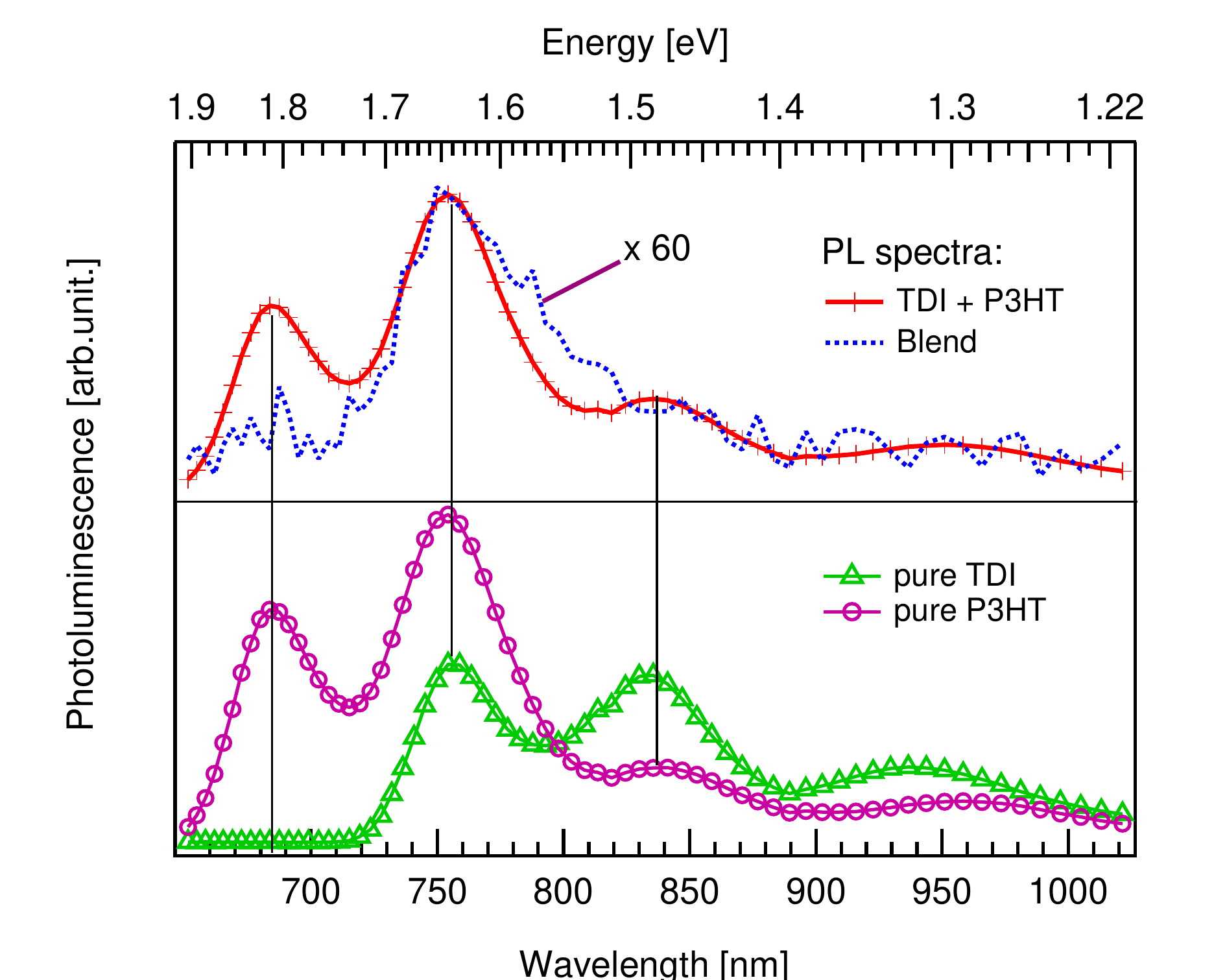}
	\caption{Sum of photoluminescences of P3HT and TDI in a 4:1 ratio (see Eq.~\ref{eqn:PLcalc}) (+), actual photoluminescence spectrum of the blend (dotted line) and photoluminescence spectra of the pristine materials (P3HT: $\bigcirc$, TDI: $\bigtriangleup$). The curves represented here were measured at 30~K, temperatures up to 300K give quite similar results in terms of quenching.
	\label{fig:PL}}
	\end{center}
\end{figure}

\subsection{Charge generation}

As shown in Fig.~\ref{fig:absorption}, the absorption spectra of TDI extends the one of P3HT. In contrast the absorption of PDI occurs in the spectral same region as the one of P3HT with main absorption peak between 530 and 535~nm.~\cite{nolde2006} We thus focus our study on TDI and use PDI mainly for comparison. We expect both molecules to have the same ability to attract the electrons from associated polymer, because they have the same LUMO level. Various blend ratios were tested, weight ratio for P3HT:TDI between 3:1 and 4:1 appeared to give the best results regarding photoluminescence quenching and PIA signal intensity as well as solar cells efficiency. The absorption spectrum of a 3:1 blend is shown in Fig.~\ref{fig:absorption}; it corresponds to the superposition of the pristine materials absorptions.

PL spectra of pristine P3HT, TDI and P3HT:TDI (ratio 4:1) blend films were recorded. Both pristine materials show characteristic PL (see inset to Fig.~\ref{fig:PL}). In order to better visualize the quenching contribution, we calculated the photoluminescence that a 4:1 P3HT:TDI blend would have in the absence of this quenching, according to the following expression:
\begin{equation}
	PL_{calc}(\lambda) = \frac{4}{5} \times PL_{P3HT}(\lambda)+ \frac{1}{5} \times PL_{TDI}(\lambda)
	\label{eqn:PLcalc}
\end{equation} 
where PL$_{P3HT}$ is the PL of the pure P3HT and PL$_{TDI}$ the one of pure TDI. This PL is shown in solid line in Fig.~\ref{fig:PL}. In contrast to the pristine materials, the blend exhibited very weak luminescence, the magnitude of the highest peak being only 1.7$\%$ of the calculated one. As a comparison, we obtained 4.1$\%$ residual luminescence with a P3HT:PCBM 1:1 blend under the same conditions (see table~\ref{tab:summary}).

\begin{figure}[t]
	\begin{center}
	\includegraphics[width=9cm]{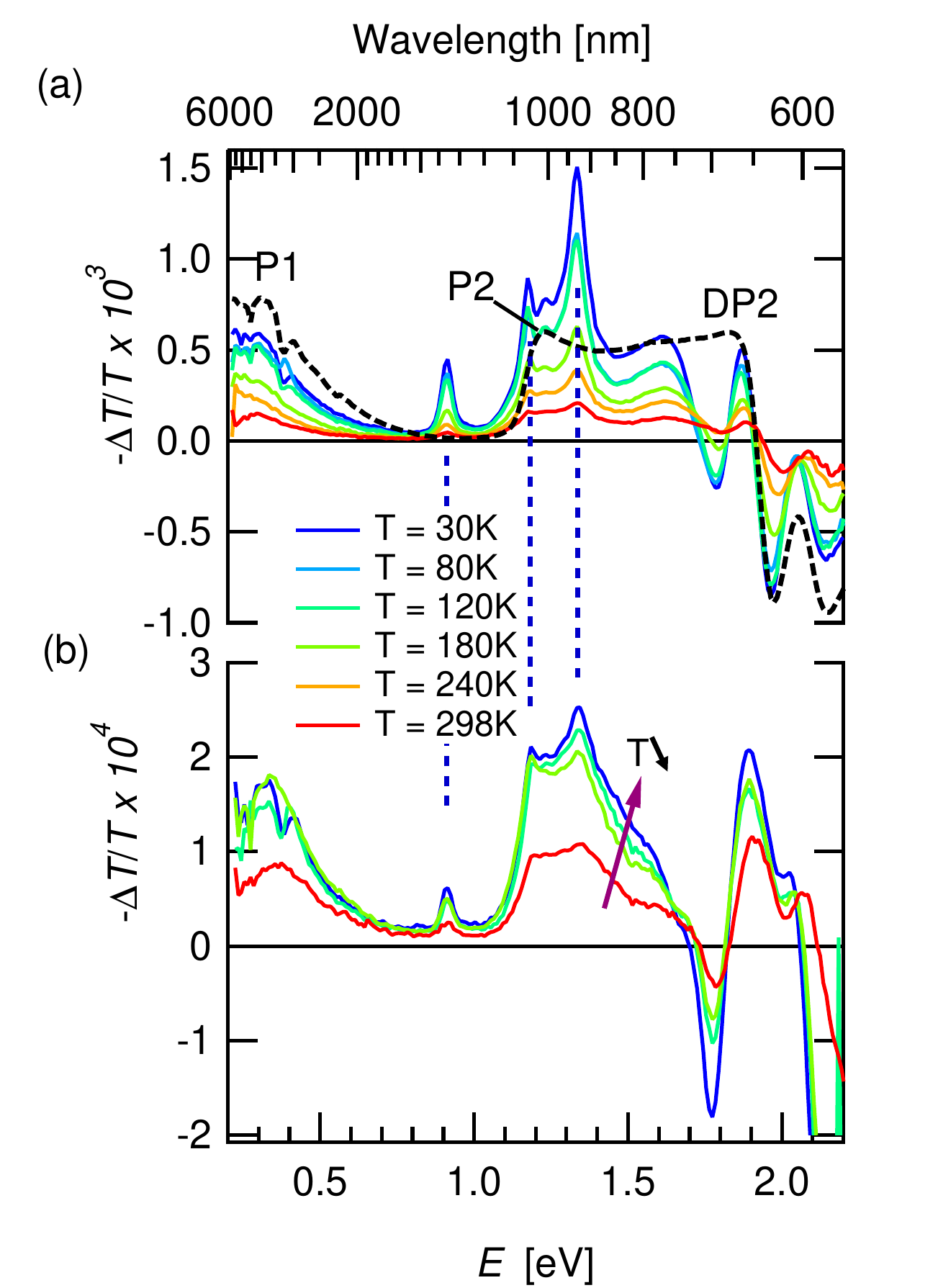}%
	\caption{(colors online) (a) Photoinduced absorption spectra of a P3HT:TDI blend at various temperatures (continuous lines) and P3HT:PCBM blend at 30K (dotted line). (b) Photoinduced absorption spectra of a MEH-PPV:TDI blend at various temperatures.}%
	\label{fig:PIATDI}
	\end{center}
\end{figure}

\begin{figure}
	\begin{center}
	\includegraphics[width=8.5cm]{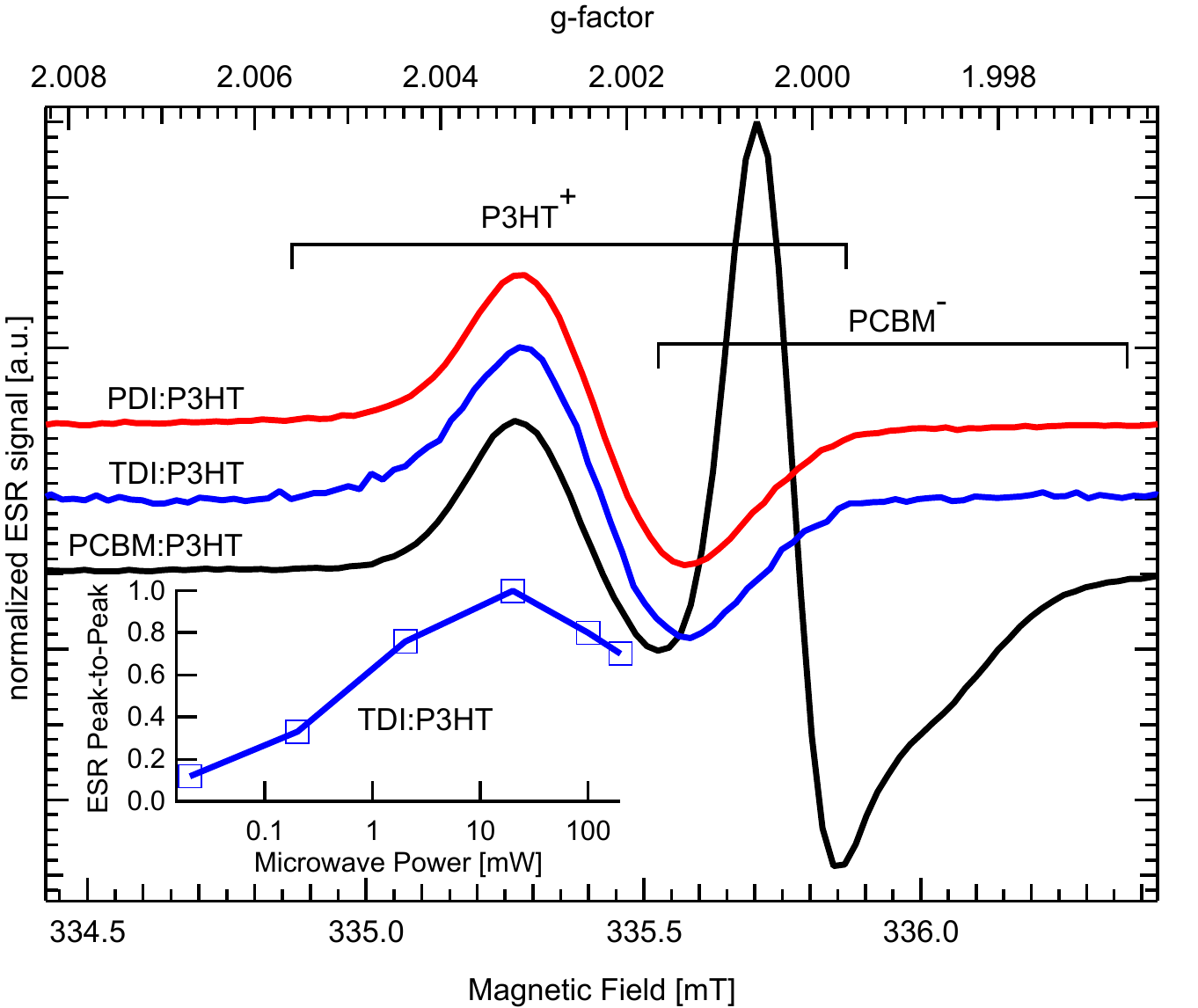}%
	\caption{Normalized ESR spectra of P3HT blended with TDI, PDI and PCBM measured at 100~K with a microwave power of 20~mW and white light illumination. The inset shows the microwave saturation behavior of the peak-to-peak signal intensity of the P3HT:TDI blend.
	\label{fig:ESR}}
	\end{center}
\end{figure}

This luminescence quenching suggests that most of the primary photo-generated excitons are converted into non-emitting excited species before they radiatively recombine, which was reported to occur within 1~ns after excitation in pure P3HT.\cite{piris2009} Moreover the shape of the remaining luminescence roughly corresponds to the calculated one which indicates that the primary excited species are indeed the same ones as in the pristine materials. Nevertheless, one can notice that the peak at 680~nm which corresponds exclusively to PL from P3HT (see inset to Fig.~\ref{fig:PL}) is more quenched than the peaks in the spectral region where both materials are luminescing. This implies that P3HT excitons are a bit more efficiently quenched than TDI excitons. This difference can be explained in terms of charge affinity of the two materials if we consider --- as will be shown below --- that this PL quenching is due to exciton dissociation into charges at the heterointerface. In case of P3HT excitons, the electron is attracted by the very electronegative TDI molecule as expected from the significant LUMO offset. In contrast, TDI excitons would be dissociated by the hole transfer into the P3HT, which is a bit less favorable, as expected from the lower HOMO offset between both materials (table~\ref{tab:summary}). Another possible reason would be that excitons in P3HT can reach the heterojunction interface more easily than those in TDI, either because of superior exciton diffusion coefficient or lower average distance to travel.

Further insight of the fate of those excitons can be obtained by studying the photo-induced absorption spectrum of the P3HT:TDI blend, which is shown in Fig.~\ref{fig:PIATDI}a. This figure clearly exhibits the characteristic absorption for P3HT localised (P1 and P2) and delocalised (DP2) polarons as reported by \"Osterbacka et al. \cite{osterbacka2000, jiang2002a} Those peaks can also be seen in other blends involving the formation of P3HT polarons such as P3HT:PCBM (dotted line shown for comparison).

In addition to those P3HT features are the 0.91~eV, 1.17~eV and 1.33~eV peaks. Similar peaks, though blue shifted by 0.4~eV, were already mentioned for PDI~\cite{gomez2007} and attributed to PDI negative radical anions (0.4~eV is also difference between the HOMO-LUMO gaps of TDI and PDI respectively, the same blue-shift can also be found in absorptions and photoluminescence).\cite{nolde2006} The position of those peaks appeared to be quite stable with temperature, as well as by varying the donor:acceptor ratio or the donor. We also observed them at the same position when using MEH-PPV as a donor (see Fig.~\ref{fig:PIATDI}b). For all those reasons, we attribute those peaks to TDI anions.

Those features, characteristic for charged species, are a strong indication that the initially generated excitons are efficiently separated into positive P3HT polarons and negative TDI anions. An additional indication is given by the absence of the peak at 1.05~eV. This peak is a very clear feature in pure P3HT (not shown) and has been ascribed to neutral excited species in pure P3HT;~\cite{osterbacka2000, jiang2002a, vanhal1999} it is totally missing in blends.

Those spectra also confirm that absorption indeed occurs in both materials as seen by the P3HT photobleaching from 1.9 to 2.2~eV (also seen in P3HT:PCBM) and one negative peak that we attribute to TDI photobleaching at 1.78~eV (corresponding to the maximum of the second absorption peak of TDI).

Fig.~\ref{fig:ESR} shows the ESR measurements which confirmed the presence of photoinduced charges in blends of TDI or PDI with P3HT. A single ESR line was observed, which exhibited resonance features such as g-factor, linewidth and saturation under high microwave power conditions (inset in Fig.~\ref{fig:ESR}) quite typical for positive P3HT polarons as in P3HT:PCBM blends \cite{marumoto2002}. However no signal of the negative TDI or PDI polaron was detected. We assume that this signal might have similar resonance features as the P3HT polaron and is superimposed by the more intense P3HT polaron spectrum. The relative signal intensities for blends of P3HT blended with TDI, PDI and PCBM were measured and are shown in Table 1.

\subsection{Solar cells}

\begin{figure}[t]
	\begin{center}
	\includegraphics[width=9cm]{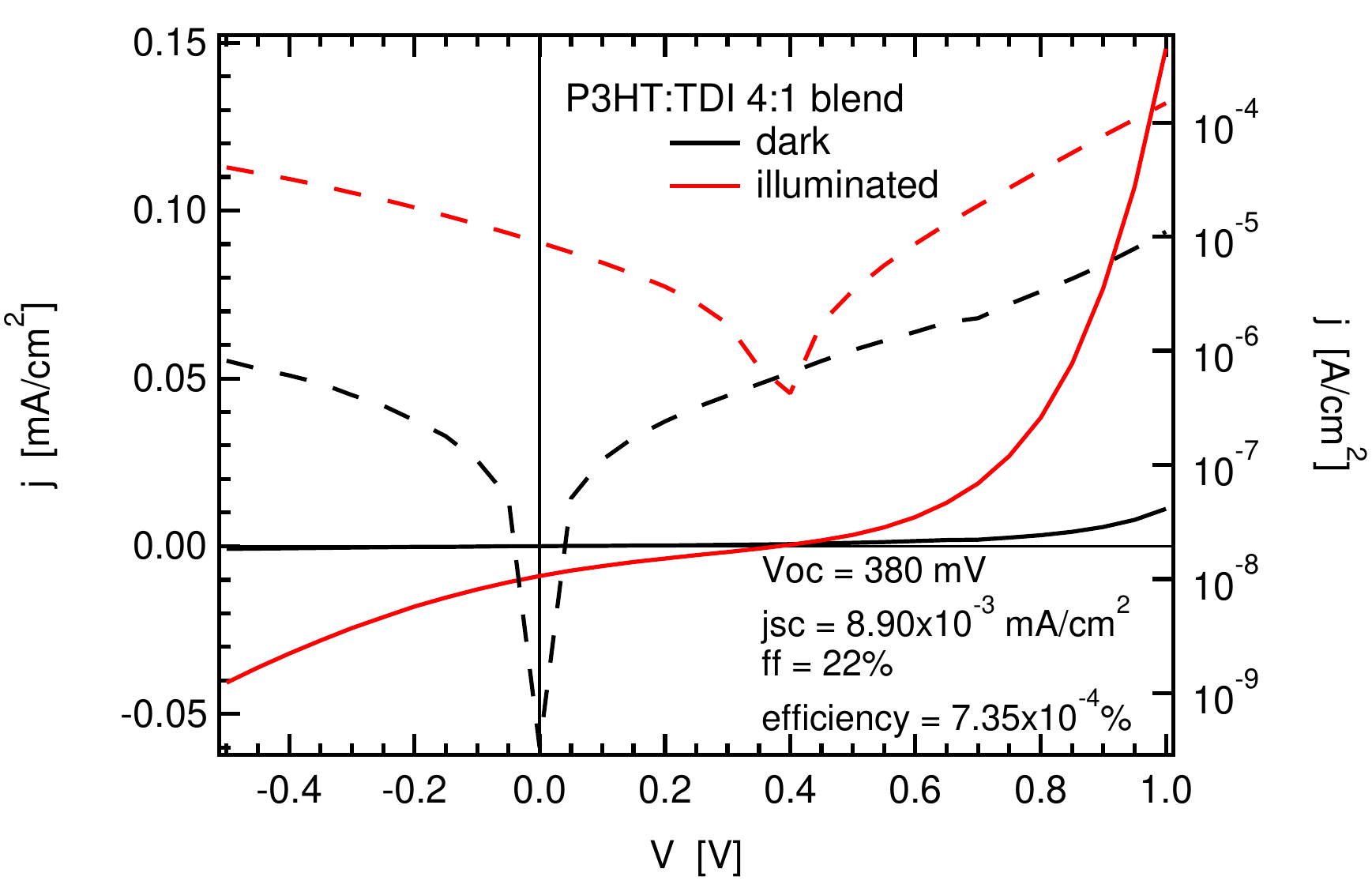}%
	\caption{Current--voltage characteristics of a P3HT:TDI solar cell with 0.5~cm$^2$ active area in dark and under halogen lamp 100~mWcm$^{-2}$ illumination represented on linear (left axis, continuous lines) and logarithmic (right axis, dotted line) current scale.%
	\label{fig:solar cell}}
	\end{center}
\end{figure}

\begin{figure}[t]
	\begin{center}
	\includegraphics[width=9cm]{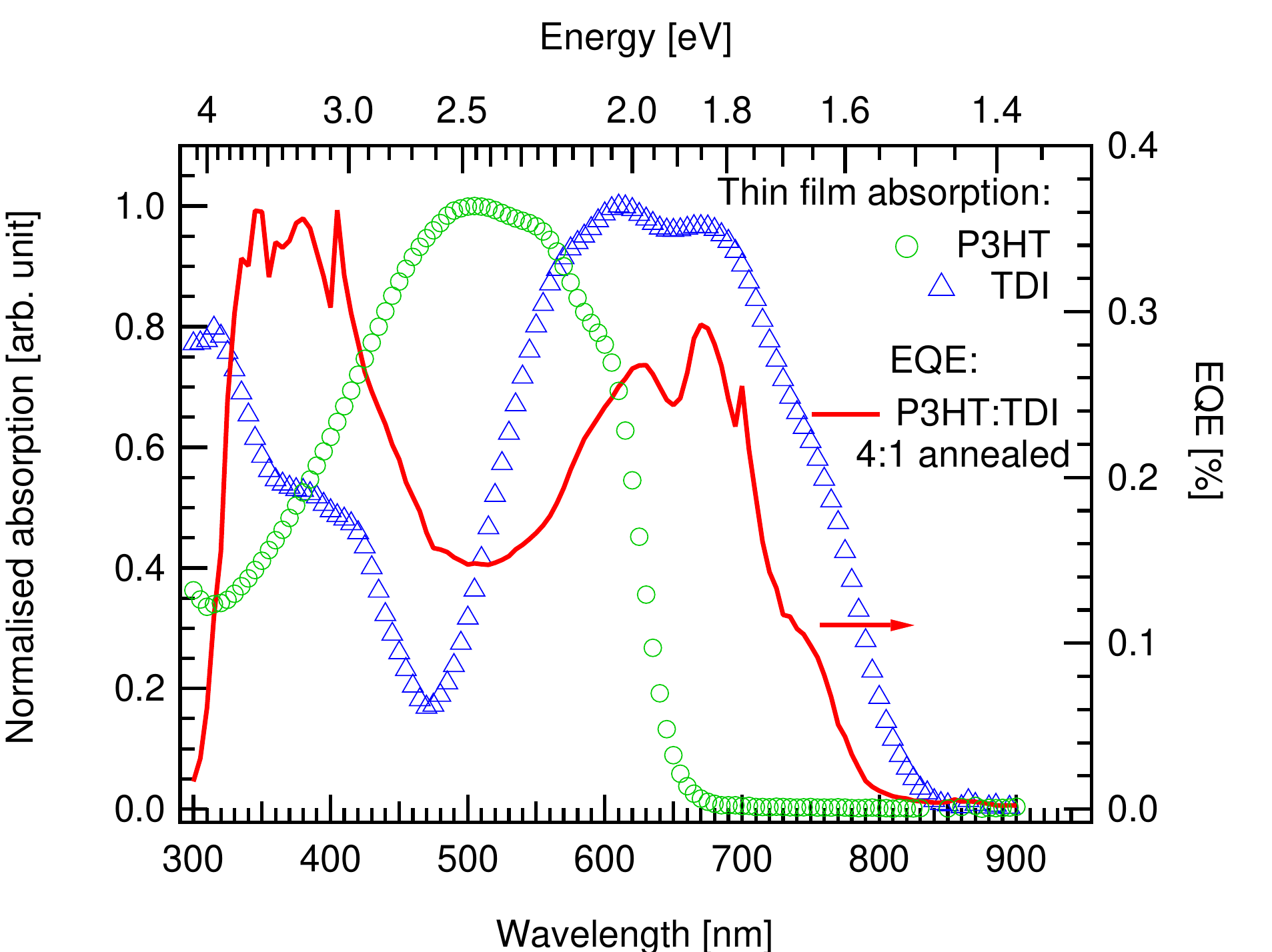}
	\caption{EQE spectrum of 4:1 P3HT:TDI blend (continuous line, left axis), normalised absorption spectra of pristine P3HT thin film ($\bigcirc$) and pristine TDI thin film ($\bigtriangleup$)
	\label{fig:EQE}}
	\end{center}
\end{figure}

Using the P3HT:TDI blend as an active layer, we were able to produce diodes. Nevertheless the photovoltaic response of the devices remained quite below our expectations. A typical current voltage characteristic is shown in Fig.~\ref{fig:solar cell}. The open circuit voltage remained comparable to what can be observed in P3HT:PCBM cells (see Table~\ref{tab:summary}), but the photocurrent was very low which contrasts quite strongly with the rather efficient charge generation observed. Additionally the photocurrent is exponentially increasing when getting away from the open circuit voltage, which results in a low fill factor and suggests a field dependent generation, and is here again in contrast with the afore mentioned good charge generation properties. 

The EQE spectrum of those solar cells is shown in Fig.~\ref{fig:EQE}. It reveals that the charges generated at 500~nm (as shown by the PIA and ESR measurements) --- and more generally in the spectral region where mainly P3HT is absorbing --- are not well extracted. In contrast in the regions where mainly or even only TDI is absorbing (above 680~nm), the EQE is higher. Indeed the EQE spectrum looks quite similar to the absorption spectrum of pure TDI with a distortion in favour of the UV region. This overweighting of the UV-regions could be due to the fact that photons of the UV region could generate excitons with enough thermal energy to overcome their binding energy and split into charges even without the assistance of an heterojunction similar to what occurs in pure P3HT diodes with photons of energy over 2.6~eV.~\cite{deibel2010}%

\subsection{Influence of the energy levels}

\ctable[
caption = Summary of the blends performances, 
label = tab:summary,
star,	
notespar,
pos = ht
]
{lcccccccccc}
{
\tnote[a] {Calculated from cyclovoltametry measurements shown in \cite{lee1999a} using the method presented in \cite{pron2010review}} plus data from \cite{li1999} and \cite{shin2007a}
\tnote[b] {Ratio providing the best efficiencies}
\tnote[c] {integrated P1 peak at room temperature, normalized to the P1 peak of the P3HT:TDI blend}
\tnote[d] {Integrated ESR signal, relatively to P3HT:TDI, at 100K}
\tnote[e] {Samples prepared from chlorobenzene solution}
}
{
	\FL					& 		 						& \multicolumn{2}{c}{Energy levels [eV]\tmark[a]}& Excit. harvest.	& \multicolumn{2}{c}{Charge separat.}& \multicolumn{4}{c}{Solar cells}							\\
\raisebox{1.5ex}{blend}	& \raisebox{1.5ex}{ratio\tmark[b]}&	$\Delta$HOMO	& 	$\Delta$LUMO	& PL quench.		& PIA\tmark[c]	& ESR\tmark[d]		& $V_0$ [V]	& $j_{sc}$ [$mA/cm^2$]	& fill fact.	& efficiency 	\\ \hline\hline
		P3HT:TDI		&		4:1						&		0.4			&		0.2			&	$\approx 98\%$	& 		1		&		1				&	0.39	&1.6 $\times 10^{-2}$	&	33$\%$		& 1.0 $\times 10^{-3} \%$	\\
		P3HT:PDI		&		3:1						&		0.4			&		0.5			&	$\approx 97\%$	& $\approx 8.8$	&		2-3				&	0.39	&2.3 $\times 10^{-2}$	&	25$\%$		& 1.4 $\times 10^{-3} \%$	\\
		PPV:TDI			&		2:1						&		0.8			&		0.4			&		$> 99\%$	& $\approx 1.5$	&		NA				&	0.74	& 		0.12			&	26$\%$		& 1.6 $\times 10^{-2} \%$	\\
		P3HT:PCBM\tmark[e]		&		1:1						&		0.4			&		0.9			&	$\approx 96\%$	& $\approx 5$	&	$\approx 10$		&	0.62 	&		10.33			&	57$\%$ 		&	3.63$\%$ 				\\ \hline
}

To study the influence of charge generation and separation we reproduced the experiments changing one of the component of the blend in order to increase the offset between the HOMO levels of the two components of the blend. This offset is indeed rather small in the P3HT:TDI blend which might leave some possibilities of energy transfer from P3HT to TDI which has a smaller band gap. Such a change should also increase the driving force for charge separation. Nevertheless it should be noted that none of the photoluminescence measurements revealed the occurrence of such energy transfer which should have caused enhanced TDI luminescence in the blend. To increase the HOMO offset, two approaches were considered: using an electron donor with higher HOMO level in a MEH-PPV:TDI blend, and using an electron acceptor with lower HOMO level in a P3HT:PDI blend (see HOMO offsets in Table~\ref{tab:summary}). The blends were studied in following ratios: 2:1 for MEH-PPV:TDI and 3:1 for P3HT:PDI. Although PIA measurements are not quantitatively comparable from one spectrum to the next, general trends can be extracted from them.

The results for these studies are summarized in Table~\ref{tab:summary}. Both blends exhibited very good harvesting of the primary excitons, characterized by a high quenching of the materials photoluminescence. In the MEH-PPV:TDI blend, the remaining photoluminescence was even below the limits of detection except at the lowest temperature (30~K), though pristine MEH-PPV has the strongest luminescence of the studied materials. The charge generation measurements (PIA and ESR) and current--voltage characteristics revealed an absence of correlation between the photophysical characteristics and device performances. Indeed, although the use of PDI instead of TDI results in a clear enhancement of charge generation (see Table.~\ref{tab:summary}), both P3HT:PDI and P3HT:TDI based solar cells exhibit a very similar performance, which is a good indication of a process preventing the charges from being extracted, independently of their concentration. In contrast, the MEH-PPV:TDI blend which exhibits smaller P1 peak (see Fig.~\ref{fig:PIATDI}b) enabled us to process solar cells with a short circuit current one order of magnitude higher than the P3HT:rylene blends.

\subsection{Charge transport properties in the P3HT:TDI blend}

\begin{figure}[t]
	\begin{center}
	\includegraphics[width=9cm]{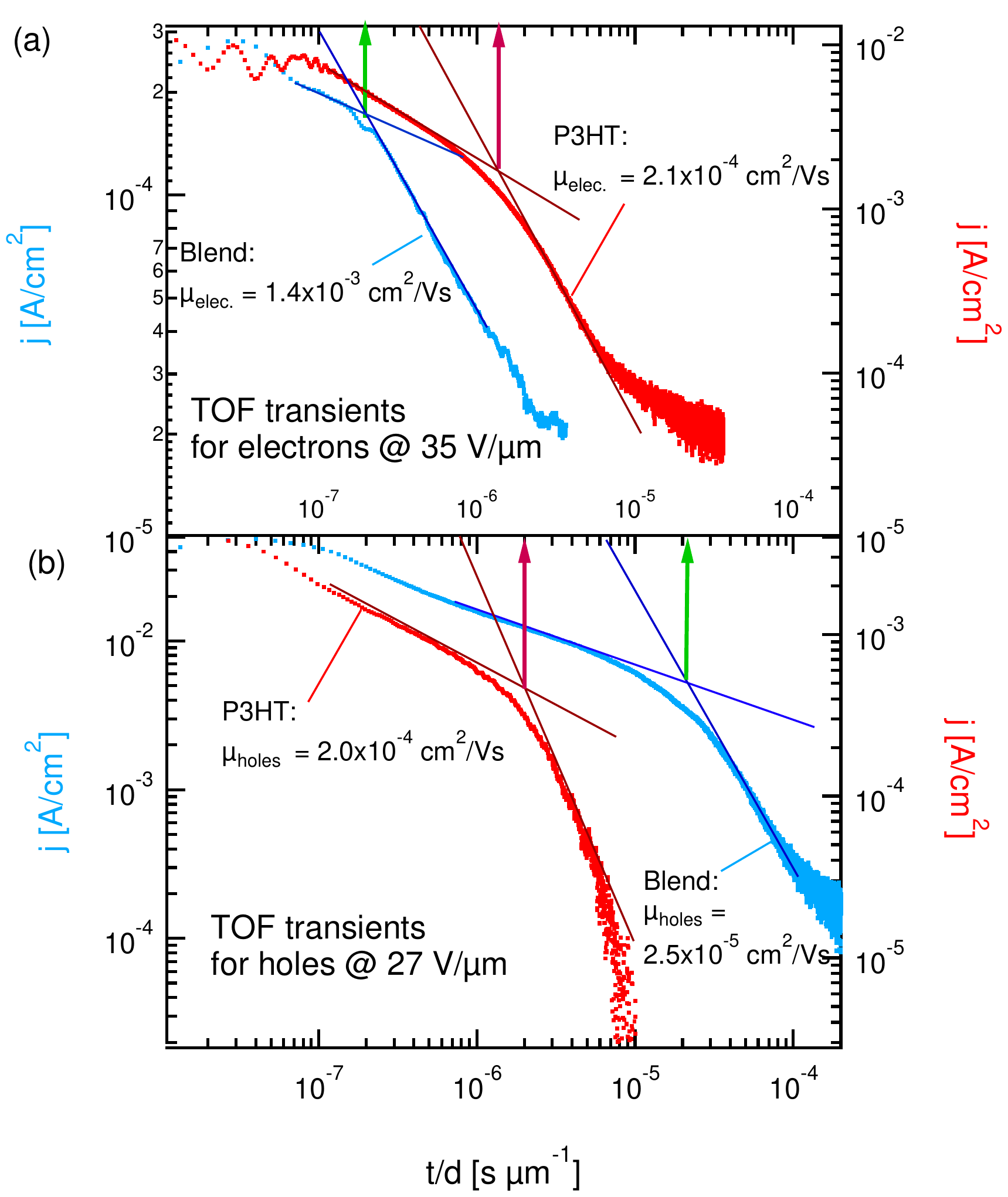}
	\caption{(Color online) (a) Electron and (b) hole current transients for the pristine P3HT and P3HT:TDI 4:1 blend. The time scale has been normalised to the thickness of the sample, the arrow thus represents the transit times $t_{tr}$ divided by the films thicknesses, which is to say the time required by charges to travel 1~$\mu$m.
	\label{fig:TOF}}
	\end{center}
\end{figure}

Possible intrinsic limitations regarding electron transport in TDI were investigated by time-of-flight photoconductivity measurements. Due to excessive roughness of pristine TDI films, no working device could be processed from them. Measurements were thus carried out on P3HT:TDI blends. They revealed an electron mobility between 1.07$\times 10^{-3}$~cm$^2$s$^{-1}$V$^{-1}$ and 2.37$\times 10^{-3}$~cm$^2$s$^{-1}$V$^{-1}$ (fig.~\ref{fig:TOF}a) for electric fields ranging from 3.3$\times 10^7$ Vm$^{-1}$ to 4.0$\times 10^7$ Vm$^{-1}$. This value is nearly one order of magnitude larger than our measurements of the electron mobility with the same method in pristine P3HT and comparable to the electron mobilities obtained in P3HT:PCBM with excess of PCBM.\cite{baumann2008} We therefore attribute it to electron transport through TDI.

The hole mobility on the same sample was found to range from 2.95$\times 10^{-5}$ cm$^2$s$^{-1}$V$^{-1}$ to 6.92$\times 10^{-5}$ cm$^2$s$^{-1}$V$^{-1}$ for fields between 6.7$\times 10^6$ Vm$^{-1}$ and 3.3$\times 10^7$ Vm$^{-1}$. As represented in Fig. ~\ref{fig:TOF}b, this is one order of magnitude lower than in pure P3HT. This is expected in blends compared to pure samples as a percolated pathway is longer than a straightforward way in a bulk, we therefore attribute this mobility to holes transported in P3HT. Such decrease has also been observed when blending P3HT with PCBM although less important\cite{baumann2008}, which suggest a less favourable morphology in the P3HT:TDI blend than in P3HT:PCBM.

\subsection{Morphology of the P3HT:TDI blend}

\begin{figure}
	\begin{center}
	\includegraphics[width=9cm]{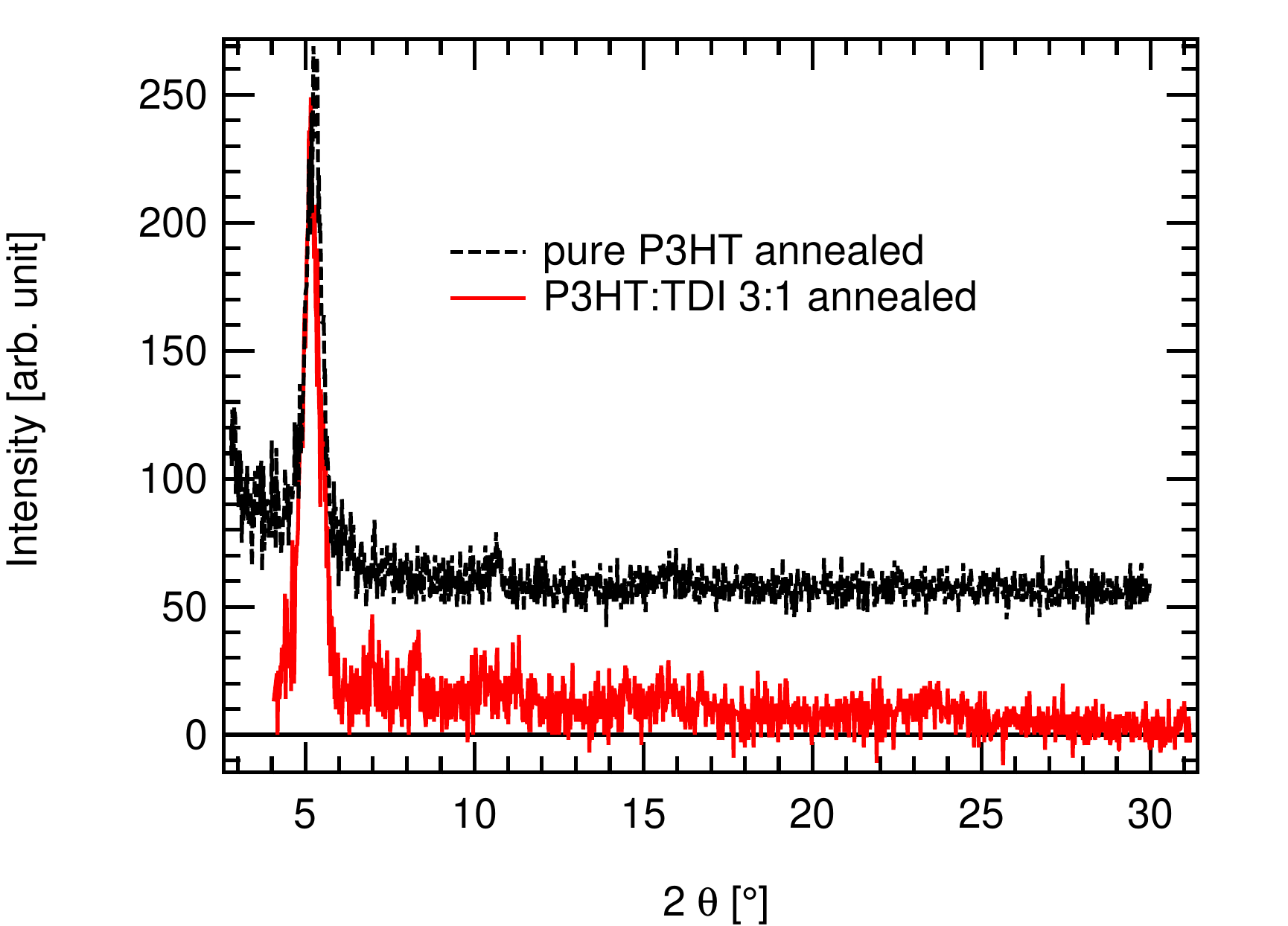}
	\caption{X-ray diffraction by a P3HT:TDI 3:1 (continuous line) sample and pristine P3HT reference in dotted line.
	\label{fig:XRD}}
	\end{center}
\end{figure}

X-ray diffraction pattern revealed in the bulk of the P3HT:TDI 3:1 sample the same feature as in the typical stacked layer of pure P3HT with a characteristic distance in the (100) direction of $\lambda /2 \sin(\theta ) = 17.1 \buildrel_{\circ}\over A$ comparable to 16.8 $\buildrel_{\circ}\over A$ found in previous works (see fig.~\ref{fig:XRD}).~\cite{prosa1992} This feature reveals an undisturbed P3HT-lamellae formation which is an indication of the presence of a pure P3HT phase in the blend.

\begin{figure}[t]
	\begin{center}
	\includegraphics[width=9cm]{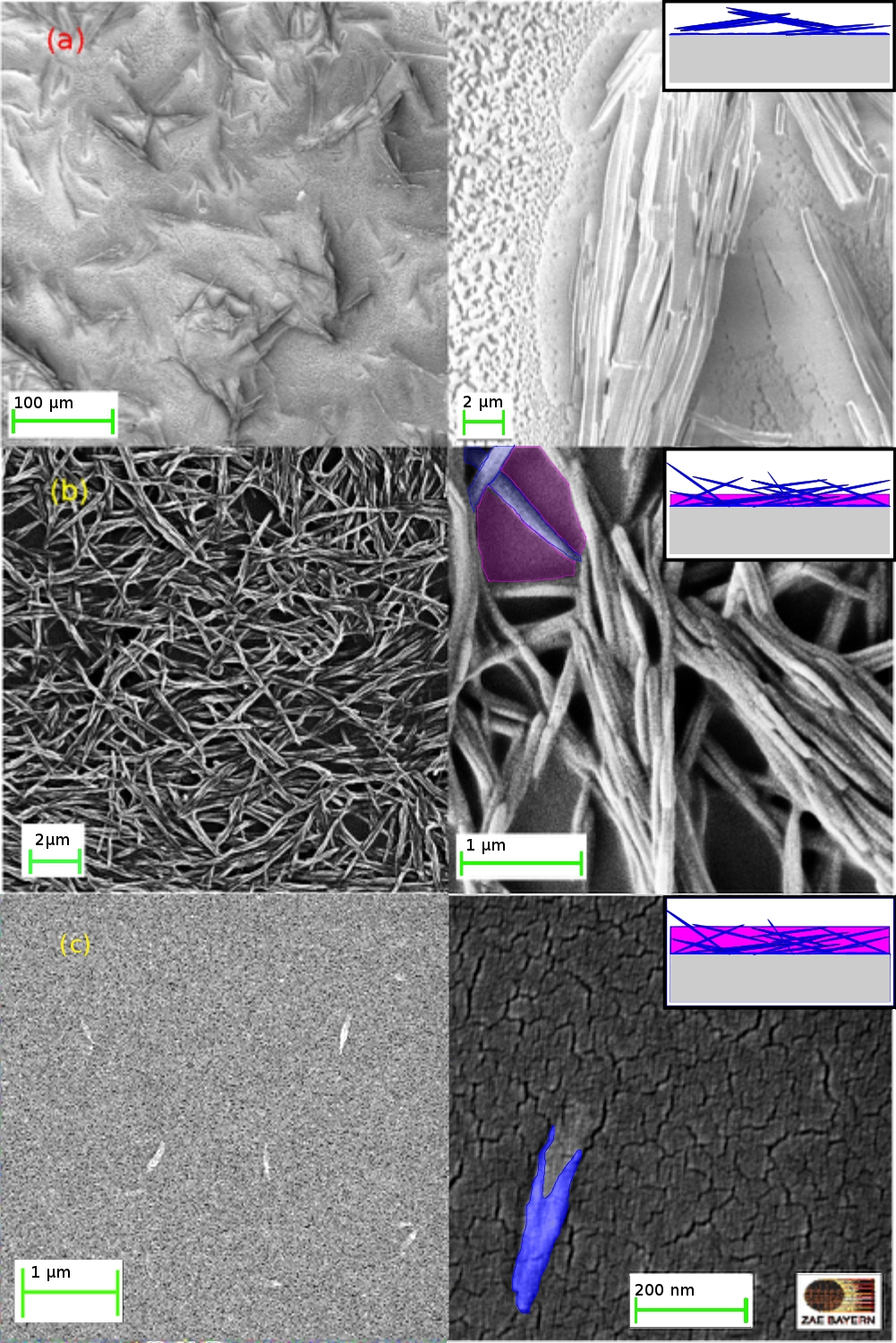}
	\caption{(color online) SEM picture of following materials spincoated on sapphire substrates: pure TDI (a) and P3HT:TDI blends with ratio 1:1 (b) and 4:1 (c). The size of the needles is visibly decreasing when increasing the ratio of P3HT. The cracks visible on the 4:1 blend are due to a thin layer of gold deposited on the organic layer for imaging. Schematics in the inset represent the possible bulk structures with blue representing TDI, fuchsia representing P3HT. 
	\label{fig:REM}}
	\end{center}
\end{figure}

Scanning electron microscopy performed on pristine TDI as well as P3HT:TDI 1:1 and 4:1 blends is represented in Fig.~\ref{fig:REM}. The images are consistent with the afore-mentioned phase separation as they exhibit two very different kind of structures. Needle-like structures are present in the pure TDI as well as in the blends and were already observed in PDI.\cite{zhou2007,li2007d} Their size is decreasing with the ratio of TDI. We therefore attribute those needles to pure TDI phases. Between the needles: a flat surface, the level of which is increasing with the ratio of P3HT to finally leave just a few needles emerge in the 4:1 ratio. We attribute this feature to the P3HT phase. It should be noted that the proportion of TDI visible at the surface of the 4:1 sample is much lower than 20 $\%$, suggesting a non homogeneous distribution of the materials. Finally a third kind of feature has to be mentioned: pure TDI also forms a flat surface on the substrate which differs from the P3HT surface by its larger roughness. Although it can be observed only in the pure TDI film, this layer could also be present in the P3HT:TDI under the P3HT. 

Those images also make clear why a ratio of at least 3:1 is necessary to obtain good photophysical properties. In the 1:1 ratio, a big part of the TDI needles are not covered by the P3HT phase, thus making it impossible for the TDI excitons to reach a donor:acceptor interface, resulting in a strong remaining TDI photoluminescence in case of a 1:1 ratio (not shown). It also explains the problem of making devices using pure material where needles height typically reaches several hundred nanometers, which is an obstacle to the evaporation of an electrode as thin as 50~nm required for time of flight measurements.

\section{Discussion}

The most remarkable observation resulting from our studies is the mismatch between the very good properties of \tdi as electron acceptor and electron transporter in P3HT:TDI blends and the poor performances of the solar cells processed from the same blend. Photo-induced absorption and electron spin resonance measurement concomitantly show that illuminating a P3HT:TDI blend with a 500~nm light results in the generation of separated charges. 

Although the charge generation at 500~nm is still lower than the one of P3HT:PCBM blend, this lower generation at a given wavelength should be partly compensated in a solar cell based on P3HT:TDI blend by the absorption over a larger region of the solar spectrum. Moreover, replacing TDI by the identically substituted PDI showed that the amount of generated charges could not be responsible for the low device performances. Indeed, by using PDI the amount of charges generated at 500~nm was multiplied by a factor between two and nine as proven by ESR and PIA measurements, although the solar cells performances remained similar showing the existence of a factor preventing those charges from being extracted of the device, independently of their concentration.

Another considered limitation is the intrinsic ability of charges to move in this material. Literature mentions crystalline PDI with local mobility as high as 0.2~cm$^2$V$^{-1}$s$^{-1}$.\cite{struijk2000} Time of flight measurements carried out in our blends revealed an electron mobility quite comparable to the one of PCBM, thus not being a limitation factor able to explain the very low current observed in our devices.

The origin of the low performance lies in the morphology of the blend. XRD measurements and SEM images give an indication that phase separation indeed occurs. The size of the phases does not seem to be an obstacle either, as we would not expect a good photoluminescence quenching in case of phases too big for the exciton diffusion length. On the opposite, too small phases would result in even better photoluminescence quenching such as the one we observe in MEH-PPV:TDI blend. Recent work~\cite{wojcik2010} suggests that too little phase separation could be an obstacle for charge separation resulting in lower PIA signal, which was also observed in MEH-PPV:TDI (see Fig~\ref{fig:PIATDI}). In spite of this, the efficiency is still higher in MEH-PPV:TDI solar cells. This suggests that in contrast with MEH-PPV:TDI solar cells where the efficiency could be limited by a factor such as the phase size, in P3HT:TDI cells, charge extraction is prevented by another factor.

The EQE spectrum of P3HT:TDI shows that very little photocurrent is originating from the P3HT absorbing regions, where the maximum of the EQE would be expected by comparison with P3HT:PCBM cells. We know that charges are generated in that spectral region because that is where the samples are excited in the PIA experiment. We propose that a thin pure TDI layer --- similar to the ones observed on pure TDI samples--- could be formed close to the substrate in the P3HT:TDI blend. This layer could efficiently prevent the holes in P3HT from reaching the anode. This assumption is consistent with the observation of a TDI layer on the substrate by SEM imageing of the pure TDI sample. This layer could also be present in the blend, even if not observable because lying under the P3HT.

The higher quantum efficiency in the spectral regions where only TDI absorbs can be explained by pristine material photovoltaic effect: the hot excitons are generated, separated and extracted within the TDI phase without the assistance of P3HT due to thermally assisted dissociation and the energetic disorder inherent to organic semiconductors. This is consistent with the SEM measurements that revealed that some TDI needles can reach the surface of the blend, thus being in contact with the anode. Photoluminescence studies also demonstrated that the contribution of TDI in the remaining luminescence in the blend is higher than the one of P3HT, proving that some excitons can remain in the TDI phase long enough for recombination to occur.

To evaluate the magnitude that charge dissociation can reach in a pure material, we processed pure P3HT solar cells. The EQE of those cells reached a maximum of 4.3$\times 10^{-3}$, similar to the P3HT:TDI solar cells whose EQE reaches a maximum of 3.7$\times 10^{-3}$. Another interesting point is the shape of the EQE spectrum: in contrast to a system in which the charge separation would be driven mainly by the heterojunction, the highest energy peaks are quite overweighted relatively to the absorption, which is a sign that the driving energy for charge separation could be the over-absorption-gap energy of the incident photons (see also~\cite{deibel2010}). 

This hypothesis is also in agreement with the low hole mobility observed in the P3HT:TDI sample. Indeed, in a TOF experiment, all the countercharges are immediately extracted. In a solar cells, holes would probably recombine when reaching the electron-rich TDI layer. In the TOF experiment this recombination can not occur, holes are thus eventually extracted but the necessity to cross the TDI layer before reaching the anode makes it much slower.

A last indication of the formation of this TDI layer was obtained by the processing of a bilayer solar cell in which a layer of P3HT was deposited under a layer of TDI thanks to the use of 2 different solvents (chlorobenzene and dichloromethane, respectively). Those cells enabled a much higher open circuit voltage (550mV), possibly indicating the open-circuit voltage is limited by space charge in the bulk-heterojunction cells. Nevertheless, due to intrinsic limitation of bilayers (rather poor charge generation), this structure did not notably allow for a better efficiency than the bulk heterojunction. We suggest that this problem could be solved by the use of devices with inverted structure (anode as top electrode).

A possible reason why this phenomenon does not appear in MEH-PPV:TDI solar cell is that in contrast with MEH-PPV, regio regular P3HT has a crystalline structure. Because of this tendency to crystallize, both P3HT and TDI crystallisations are in competition thus applying some mechanical constraints on each other which may push some TDI under the P3HT containing part until forming this hole blocking layer. Another possibility is that MEH-PPV and TDI do not have fully separated phases, which would explain the low yield of charge generation observed.

\section{Conclusion}

In TDI, we found an organic material exhibiting very interesting properties regarding its use as electron acceptor in blend with P3HT. Good charge generation and transport together with an absorption band complementary to the one of P3HT were demonstrated, promising good photovoltaic properties to the P3HT:TDI blend. Nevertheless, the solar cells produced using this blend as active layer resulted in low performances. In agreement with our morphology investigations, we present two explanations. In the case of MEH-PPV:TDI, we propose that the photocurrent is severely limited by the too small phase separation. In P3HT:TDI solar cells, the formation of a thin pure TDI layer along the ITO electrode seems prevent hole extraction. Recent work by Kamm et al.~\cite{kamm2011} on PDI derivatives showed that replacing N-subsituted by core-substituted rylene derivatives in blends with P3HT enabled to prevent rylenes aggregation and improve solar cell performances. As a result one could expect that core-substituted TDI derivative also exhibit a morphology more compatible with solar cell applications, thus enabling the very good potential of those acceptors to be used in good performing solar cells collecting a wider range of the solar spectra.

\section{Acknowledgements}

This work was supported by the European Commission through the Human Potential Program (Marie-Curie RTN "SolarNType" Contract No. MRTN-CT-2006-035533; website: www.solarntype.eu). V.D.s work at the ZAE Bayern is financed by the Bavarian Ministry of Economic Affairs, Infrastructure, Transport and Technology. C.D. gratefully acknowledges the support of the Bavarian Academy of Sciences and Humanities.


\bibliographystyle{unsrt}
\bibliography{Papers}

\end{document}